# Critical Role of Disorder for Superconductivity in the Series of Epitaxial Ti(O,N) Films

**Authors:** Fengmiao Li[1,2], Oliver Dicks[2], Myung-Geun Han[3], Solveig Aamlid[2], Giorgio Levy[1,2], Ronny Sutarto[4], Chong Liu[1,2], Hsiang-Hsi Kung[1,2], Oleksandr Foyevstov[1,2], Simon Godin[1,2], Bruce A. Davidson[1,2], Andrea Damascelli[1,2], Yimei Zhu[3], Christoph Heil[5], Ilya Elfimov[1,2], George A. Sawatzky[1,2], Ke Zou[1,2]

**Affiliations:**

[1]Department of Physics & Astronomy, University of British Columbia, Vancouver, British Columbia, V6T 1Z1 Canada.

[2]Quantum Matter Institute, University of British Columbia, Vancouver, British Columbia, V6T 1Z4 Canada.

[3]Condensed Matter Physics & Materials Science Department, Brookhaven National Laboratory, Upton, New York, 11973 USA.

[4]Canadian Light Source, Saskatoon, Saskatchewan, S7N 2V3 Canada.

[5]Institute of Theoretical and Computational Physics, Graz University of Technology, NAWI Graz, 8010 Graz, Austria

**Abstract:** Realizing experimental control of superconductivity is of paramount importance to advancing both basic research and technological applications. Disorder, generally existing in most superconductors, intricately interacts with Cooper pairs and also impacts the performance of quantum devices. In this paper, we report the study of a series of Ti(O,N) crystalline films prepared via molecular beam epitaxy (MBE). We discover that substituting nitrogen (N) for oxygen (O) in TiO, namely TiO(N), considerably increases the normal-state conductivity and the superconducting transition temperature $T_c$. The $T_c$ of TiO(N) falling between those of TiO (~0.5 K) and TiN (~6 K) is contrary to their comparable $T_c$ predicted by the Migdal Eliasberg theory. It is found that their resistivity vs temperature obeys the "Mooij rule", known as the characteristic of metallic glasses. Density functional theory (DFT) calculations demonstrate that strong disorder severely scatters the Bloch electron waves at nonzero momenta, which consequently weakens electron-phonon coupling in TiO(N).

**Introduction:** The Bardeen–Cooper–Schrieffer (BCS) theory, which requires bosons such as phonons as the medium for pairing two distant electrons, serves as one of the prevailing microscopic mechanisms for superconductivity (*1*). However, disorder, which influences short-range microscopic interactions and local symmetry (*2*), coexists with superconductivity in many materials. The profound effects of disorder on electronic systems, along with the bad-metal characteristics of unconventional superconductors like cuprates (*3*) and recently-revisited nickelates (*4*), have motivated numerous studies on the interplay between disorder and superconductivity. Experimentally, the response of superconductivity to variations in disorder can provide valuable information that deepens our understanding of superconductivity and helps to unravel some of the mysteries surrounding the electron-pairing mechanism.

Conventional superconductivity, arising from the attractive interaction between two distant electrons with opposite momenta $\vec{k}$ and $-\vec{k}$, exhibits robustness against a moderate level of disorder (*5*). In contrast, many superconductors (*6*–*8*) have shown increased sensitivity to disorder. However, the major challenge is the lack of techniques capable of directly probing disorder at the nanometer scale; many microscopic studies utilizing scanning tunneling microscopy (STM) are limited to surface measurements (*9*, *10*), which may not accurately reflect the bulk properties. A generally implemented method for evaluating the disorder strength in the bulk phase, which we will also adopt in this work, is by analyzing the electron mean free path in electrical transport experiments (*11*). This method measures cumulative effects of all electron scattering processes, providing a qualitative assessment of the overall strength of various types of disorder.

The interplay between disorder and superconductivity can vary significantly across different systems. It's worthy to mention that some of key material parameters, such as the degree of crystallinity, often changes as disorder increases. Simultaneous variations in many physical parameters complicate efforts to elucidate the relationship between disorder and superconductivity. Electron-phonon coupling that binds two electrons together in elemental superconductors is still valid in many oxide superconductors such as hole-doped $BaBiO_3$ (*12*). Chemical substitution of potassium (K) replacing barium (Ba) in bismuthates, which introduces holes necessary for inducing superconductivity, causes irregular potential with an effective interaction range longer than inter-atomic distances. In strongly-correlated systems, many-body entanglements governed by quantum mechanics dominate low-energy physics with higher sensitivity to disorder. After decades of synthesis efforts yielding high-quality crystals of cuprates, unconventional superconductors are known to be inherently disordered in general (*3*). It underscores the importance of short-range inhomogeneity in charge and magnetism, which possibly play an important role in the electron pairing process.

Recently-synthesized TiO epitaxial films exhibit a $T_c$ ~0.5 K and a short electron mean free path at elevated temperatures (*13*); its nitride counterparts, titanium nitrides (TiN), which demonstrate to have good metallic properties, are superconducting with one order of magnitude higher $T_c$ of ~6 K (*14*). Superconductivity was also reported in titanium oxynitrides but in the form of polycrystals with large uncertainty particularly concerning material uniformity (*15*, *16*). This raises interesting questions about whether Ti oxynitrides are intrinsically superconducting and what the extent of the disorder presents. The isostructural TiO and TiN have close lattice parameters, and, in addition, their physical property can be understood in the framework of independent particles without many-body correlations (*13*). Thus, the Ti(O,N) series, ranging from TiO and Ti oxynitride to TiN, could serve as promising material platforms for studying the relationship between disorder and superconductivity.

In this study, we have developed a nitric oxide (NO) molecules-assisted MBE growth method to prepare the intermediate crystalline phase of TiO and TiN, *i.e.*, Ti oxynitrides. Our findings show that the substitution of N for O in TiO enhances superconductivity and increases the electron mean free path at the normal state. We observe that the conductivity as a function of temperature in Ti(O,N) materials at higher temperatures resembles that of metal alloys, following the well-known "Mooij rule" (*17*) in disordered materials. The gradual substitution of N for O increases chemical disorder within anion sublattices while simultaneously reducing intrinsic disorder present in TiO. This results in a complex interplay between two types of disorder, both of which affect superconductivity.

**MBE epitaxial film growth and superconductivity property.** The growth of Ti oxynitride TiO(N) films was performed in a Veeco GENxplor MBE chamber with the base pressure better than $1\times10^{-10}$ Torr. The MgO (001) substrates with similar lattice parameters are used as crystalline templates for epitaxial film growth. The NO gaseous environment during Ti deposition serves as the oxidation agent and sources for both N and O with the ratio exactly equal to 1. X-ray diffraction (XRD) 2θ-ω scans (Fig. 1A & C) show sharp (002) diffraction peaks close to the substrate ones, demonstrating the epitaxial growth of single-crystalline TiO and Ti oxynitride (TiO(N)) films. The out-of-plane lattice constants of TiO and TiO(N) are calculated to be 4.16 Å and 4.21 Å respectively, implying that the lattice volume expands as more N incorporates (*13*, *14*). The full width at half maximum (FWHM) of rocking curves is nearly equal at ~ 0.029° for TiO and ~0.026° for TiO(N), suggesting similar crystalline quality of the two films. *In-situ* electron diffractions and *ex-situ* reciprocal space mapping in XRD confirm the coherent growth of high-quality crystalline films (Fig. S1). The Ti and Mg atom columns near the interface are clearly resolved in the high-angle annular dark-field (HAADF) scanning transmission electron microscopy (STEM) image (Insets in Fig. 1A & C).

As shown in Fig. 1B & D, TiO and TiO(N) epitaxial films become superconducting at low temperatures. The resistance of the TiO(N) film decreases by half at $T_{c,\,mid}$ ~ 2.5 K, one order of magnitude higher than $T_{c,\,mid}$ ~ 0.4 K of TiO. As out-of-the-plane magnetic field increases, the superconducting transitions move to lower temperatures and eventually disappear at ~1.0 Tesla for both systems at the temperatures achievable in our measurement systems. The calculated Landau-Ginzburg superconducting coherence lengths are ~22 nm for TiO and ~16 nm for TiO(N) (*18*), respectively, (see detailed R(B) analysis in Fig. S2). The comparable coherent lengths in TiO and TiO(N) films suggest that the type of interaction responsible for their electron pairing is probably similar. However, the critical field of pure TiN is much smaller at 0.21T, thus giving a longer coherence length ~40 nm (*14*). This hints that some key parameter of TiN relevant to superconductivity is distinct.

**Spectroscopy characterization of Ti(O, N) films.** In order to investigate the chemical composition and low-energy electronic structures of prepared TiO(N) films, we have employed complementary element-specific spectroscopies with the bulk and surface sensitivity, including x-ray absorption spectroscopy (XAS) (Fig. 2A), x-ray photoelectron spectroscopy (XPS) (Fig. S3), and electron energy loss spectra (EELS) in STEM (Fig. S4).

Although the NO molecule naturally provides an equal number of N and O for the film growth, the higher reactivity of oxygen with Ti results in a disproportion of the anion incorporation. Analysis of the spectral weights in EELS (Fig. S4) reveals the element ratio of Ti, O and N of approximately10:6:4, with an error margin of ~±10%, corresponding to the chemical formula $TiO_{0.6}N_{0.4}$. We have consistently reproduced multiple samples with the same N and O ratio, experimentally evidencing thermodynamically stable synthesis. Additionally, atomic ordering is not observed in our STEM and XRD experiments, indicating that N and O, both occupying anion sites in the rock-salt lattice, are probably distributed randomly.

The XAS at the K edge of O & N in Fig. 2A is due to the electron excitation from the core-level 1s orbital to unoccupied 2p states, which also carries important information about Ti 3d, 4s and 4p states because of the hybridization between Ti and O or N. The separation between the first two peaks, which corresponds to the splitting of Ti 3d $t_{2g}$ and $e_g$, is greater in the N XAS than in the O XAS, indicating N exerts a stronger crystal field than O or the hybridization of Ti-N bonds is stronger. The N K edge XAS shows a good agreement with the DFT-calculated partial density of states with the p orbital character (Fig.S5 A&B). The Ti XAS at the $L_{2,3}$ edge (Fig. 2A) exhibits two primary peaks due to spin-orbit splitting of the Ti 2p core level. An additional pronounced peak, as indicated by the arrow, is attributed to the presence of $Ti^{3+}$ due to the N hole doping. Unfortunately, owing to the poor energy resolution, the $Ti^{3+}$ peak is not clearly observed in EELS (Fig. S4).

*In-situ* ultraviolet photoemission spectroscopy (UPS) in Fig. 2B measures occupied states below the chemical potential. As demonstrated by DFT calculations (Fig. S5 C&D), the states near the Fermi level are primarily the $t_{2g}$ states of Ti 3d, while the spectral weight observed at higher binding energies, between 4 eV and 8 eV, corresponds to the excitation of occupied O and N 2p states, respectively. The abrupt cutoff at the Fermi level demonstrates the metallic properties of Ti oxynitride films.

In short, the resemblance of our XAS results as well as the valence-state measurements in TiO, TiO(N) and TiN (*13*, *19*) suggest that the low-energy physics in these materials originates from $t_{2g}$ bands of Ti with only the variation of electron filling .

**Electrical transport properties and disorder.** In the study of high-$T_c$ superconductivity, insights obtained from the investigation of normal-state interactions have provided valuable clues for comprehending the microscopic mechanism of electron pairing. This context prompts us to look into the evolution of normal-state electrical transport in the series of Ti(O,N) epitaxial films (Fig. 3A-C). Interestingly, a large variation in the resistivity from TiO, $TiO_{0.6}N_{0.4}$ and TiN is observed. Despite the mixing of N and O, the $TiO_{0.6}N_{0.4}$ film exhibits a normal-state resistivity of ~100 μΩ·cm, much lower than that of TiO films at ~300 μΩ·cm (*13*). Following the trend of resistivity variation as more N is added, the MBE-grown TiN film has an exceptionally low resistivity of only several micro-ohm centimeters (*14*). The mean free paths for TiO, $TiO_{0.6}N_{0.4}$ and TiN are, respectively, ~2 Å, ~6 Å, and ~100 nm at the temperature just above their $T_c$ (*20*).

The resistivity versus temperature $\rho_{xx}(T)$ of $TiO_{0.6}N_{0.4}$ films shows a weak temperature dependence with a positive derivative, whereas the TiO resistivity exhibits a negative derivative with the temperature which indicates a bad metal or bad semiconducting property. This reveals that N chemical substitution transforms semiconducting TiO into metallic Ti oxynitride. Additionally, a small upturn beginning at~ 60 K at low temperatures (the inset in Fig. 3B) is observed. Instead, the TiN resistivity drops quickly from ~10 μΩ·cm at 300 K to ~ 1 μΩ·cm at low temperatures, manifesting that TiN is a good metal. The transformation from high-resistivity "bad" metals to good metals is realized by controlling N substitution, a phenomenon that cannot be simply explained by the Drude model and conventional band theory of chemically doped materials. We discover that the evolution of $\rho_{xx}(T)$ across the series of Ti(O,N) films, including the low-temperature upturn in Ti oxynitrides (Fig. 3B), has been observed in disordered transition-metal alloys (*21*, *22*). It closely follows the "Mooij rule" (*17*), which describes a transition from a positive to a negative derivative of $\rho_{xx}(T)$ when the electron mean free path exceeds inter-atomic distances, corresponding to the resistivity of about 150 μΩ·cm.

Experimental results of $T_c$ and normal-state resistivity, as shown in in Fig. 3D, suggest an intimate interplay between disorder and superconductivity. Importantly, both physical properties are tunable over a large range. As shown in the inset of Fig. 3D, the TiO resistance at 2 K increases at low fields due to the suppression of the superconducting states, indicative of preformed Cooper pairs in TiO from strong disorder (*23*), then decreases when the field passes over ~4 T. In contrast, the $TiO_{0.6}N_{0.4}$ films possess negligible magnetoresistance. In addition, the disorder, resulting in electron localization, may be responsible to the much lower TiO carrier density than in TiO(N) (Fig. 3E), which was observed in metal alloys (*2*).

**Electronic and phonon structures of disordered Ti(O,N).** The persistent superconductivity exhibited across materials in the Ti(O,N) series is captivating. In the weak coupling theory of superconductivity, $T_c$ is proportional to the phonon frequency and the density of states (DOS) near the Fermi level (*24*). Our DFT calculations show that TiO has a higher DOS near the Fermi level than TiN (Fig. S6), whilst having phonon frequencies of similar magnitude. Solving the semi-empirical Allen-Dynes McMillan formula gives a $T_c$ of about 12 K for TiO and about 9 K for TiN. More precise calculations of superconducting properties employing the Eliashberg equations, as implemented in the EPW code (*25*), predict TiO to have a $T_c$ in the range of 15-16 K, slightly higher than TiN (14-15 K). The notable gap between the theoretical prediction and experimental observation implies that there must be additional important interaction at play which has been neglected in our theoretical calculations implemented so far.

As we discussed in the previous section, disorder also plays a primary role in determining electrical property in the normal state. Here, we will discuss the influence of disorder on the dispersions of both band and phonon. We generated a supercell containing 64 Ti, 39 O and 25 N with a nominal stoichiometry $TiO_{0.61}N_{0.39}$, featuring quasi-randomly distributed N and O (Fig. 4A). Notably, a significant difference is observed in the calculated phonon structures between the bulk TiO and TiN with Rocksalt structures. As shown in Fig. S7, the phonon dispersions obtained for TiN are normal, in agreement with previous calculations (*26*), whereas several branches of TiO phonons exhibit negative frequencies at some momenta. This strongly suggests that the previously observed electron scattering in TiO results from its inherently unstable nature of TiO (*27*). The reduced disorder in $TiO_{0.6}N_{0.4}$ is because of N substitution, exhibiting the absence of phonon branches with negative frequencies (Fig. 4B). It demonstrates that N substitution in TiO stabilizes the Rocksalt structure.

In the bulk TiO and TiN, the dispersions of the Ti $t_{2g}$ bands within energies between 2 eV and -4 eV are similar (Fig. S8), and the observed small differences are due to the differential bonding of Ti to O and N. It is noticeable that, because of N hole doping, the Fermi level progressively moves to lower energies. The unfolded band structure of Ti oxynitrides in Fig. 4C shows that, close to the boundaries of Brillouin Zone (BZ), the band dispersions of Ti 3d $t_{2g}$ and anion (N & O) 2p are significantly smeared out. The randomly-distributed N and O removes the translational symmetry of not only O but also Ti bands, thereby resulting in decoherence in the electron Bloch waves. However, the coherence of electronic states remains well-preserved at the BZ center. It is interpreted as the fact that, , the disorder effect is averaged out over a long-range domain, not visible at the coherence of long-wavelength electrons.

A similar momentum dependence of decoherence is observed in the calculated phonon dispersion, as illustrated in Fig. 4B. While the precise nature and exact origin of the disorder in TiO require further investigation, it is certain that disorder induces similar momentum-dependent phase decoherence, but with much stronger effects. The $TiN_{0.4}O_{0.6}$ film houses considerably less disorder and thus, chemical disorder is regarded as the primary source of electron scatterings. The pure TiN, characterized as a "clean" material, has minimal disorder that exerts little disruption on the Bloch waves. Importantly, most electronic states residing near the Fermi level have a smaller $k_f$, which, in hence, are robust against disorder.

The strong disorder in TiO and $TiO_{0.6}N_{0.4}$ severely disrupts the Bloch wavefunction of electrons and phonons (*2*), suppressing the $T_c$, and thereby, leaving the Anderson theorem invalid. The Ti oxynitride, $TiO_{0.6}N_{0.4}$, although carrying extra anion intermixing, is less disordered than TiO, exhibiting an intermediate $T_c$. In the case of TiN with minimum disorder, its electronic states throughout the BZ retain their coherence, leading to the superconductivity with the highest $T_c$ and a longest coherence length. Disorder likely plays a decisive role in defining the interaction distance of paired electrons, because, in strongly-disordered systems, short-range interactions are strengthened, which could effectively reduce the coherence length.

**Concluding remarks.** In this study, we have discovered that the Ti oxynitrides, in the form of epitaxial crystalline films grown via MBE, exhibit superconductivity with $T_c \sim$ 2.6 K. Intrinsic disorder in TiO strongly reduces the coherence of the dispersive phonon and electronic structures, with the exception of those electronic states located in close proximity to the BZ center. Substituting N for O moves the Fermi level, changes the shape of Fermi surface and, more importantly, decreases the amount of disorder evidenced by electrical transport measurements. Our results emphasize that the momentum-dependent disorder effect should be considered in the future quantitative study of superconductivity in disordered systems. Our work provides material platforms for investigating the interplay between disorder and paired electrons in a large domain. The exceptional controllability of superconductivity and disorder in a range of Ti(O,N) epitaxial films are highly desired for the future advancement of superconducting circuits and single-photon detectors in quantum computers (*28–30*).

**Methods**

MBE film growth: The MgO (001) substrate was annealed at 700 ºC in ultra-high vacuum (UHV) for 1 hour before the film growth. In order to maintain the growth of high-quality films, the NO pressure in the growth chamber was kept at $\sim 1.0\times10^{-8}$ Torr within a small window of pressure variation by adjusting a leak valve. the film growth rate was determined by the Ti flux at ~1 layer/min measured using a quartz crystal microbalance (QCM). A higher or lower NO growth pressure caused undesired phases, such as $Ti_2O_3$ and Ti metal. A protective layer of Germanium (Ge) film with a thickness of ~10 nm was deposited at room temperature to protect the sample for *ex-situ* measurements.

Electrical transport measurement: Temperature-dependent resistivity from 300 K to 2 K were performed with Van der Pauw geometry in a Quantum Design physical property measurement system (PPMS). Indium metal was used as the contact electrode.

X-ray diffraction measurement: The XRD was measured using x-ray single crystal diffractometer, Bruker D8 DISCOVER at the room temperature.

Spectroscopy measurement: Room-temperature UPS measurement was done at the UBC ARPES lab. A vacuum suitcase with the vacuum better than $1\times10^{-11}$ Torr was used for transferring samples from the MBE growth chamber to the characterization chamber. The photon energy was at 21.2 eV.

Fluorescence yield of XAS was collected at 10° angle away from the sample surface with the incoming x-ray beam at 30° angle relative to the sample surface. *Ex-situ* XAS and XPS measurements were performed at Resonant Elastic and Inelastic X-Ray Scattering (REIXS) beamline of Canadian Light Source. The XPS Al x-ray source was monochromatized at 1486.6 eV with an energy resolution of ~0.4 eV calibrated at the Ag Fermi level.

STEM experiments: For STEM, Ge-capped Ti oxynitride films on MgO (001) substrates were cross-sectioned by focused ion beam technique using 5 keV $Ga^+$ ions to minimize ion beam–induced damages. A JEOL ARM 200CF equipped with a cold field emission gun and double spherical aberration correction at Brookhaven National Laboratory operated at 200 kV was used for HAADF imaging with detection angles ranging from 68 to 280 mrad. For EELS, a Gatan Quantum ER spectrometer was used with dispersion (0.1 eV/channel) and ~0.8 eV energy resolution. The convergent and collection semiangles were, respectively, ~10 and ~5 mrad.

<u>DFT calculations</u>: DFT calculations were performed within the Projector Augmented Wave method and the Perdew-Burke-Ernzerhof functional (*31*), as implemented in the Quantum Espresso package (*32*, *33*). The disordered supercell for Ti oxynitride calculations was made using the mcsqs code from the Alloy Theoretic Automated Toolkit (ATAT) (*34*). The code uses a Monte Carlo algorithm to generate Special Quasi-random Structures (SQS). For the SQS cell, a supercell (4×4×4 of the primitive unitcell) with 25 N atoms and 39 O atoms was initialized, and the correlations of anion pairs up to 6 Å, triples up to 3 Å, and quadruples up to 3 Å were optimized until the correlation mismatch was less than 0.01 for all clusters. The supercell bandstructure and phonon dispersions were unfolded using the unfold.x code (*35*). Crystal structures were plotted using the VESTA software (*36*).

We employed the EPW code package (*25*, *37–39*) for the Wannier interpolation of the electron-phonon matrix elements onto dense k- and q-grids and the subsequent self-consistent solution of the isotropic Migdal-Eliashberg equations. In particular, we used coarse 6×6×6 k- and q-grids and interpolated onto fine 24×24×24 reciprocal grids, set a Matsubara frequency cutoff of 1eV, included electronic states within ±1 eV around the Fermi energy, and chose a standard value for the Morel-Anderson pseudopotential of mu*=0.10. In a stoichiometric and primitive unitcell TiO, where imaginary phonon frequencies appear when performing phonon calculations in the harmonic approximation, all electron-phonon matrix elements corresponding to such harmonically unstable phonon modes have been set to zero.

**Acknowledgements**

This research was undertaken thanks in part to funding from the Max Planck-UBC-

UTokyo Centre for Quantum Materials and the Canada First Research Excellence Fund, Quantum Materials and Future Technologies Program. The work at UBC was also supported by the Natural Sciences and Engineering Research Council of Canada (NSERC), British Columbia Knowledge Development fund (BCKDF), Canada Foundation for Innovation (CFI), Canada Research Chair Program (A.D.), and CIFAR Quantum Materials Program (A.D.). XAS and XPS characterizations were performed at the Canadian Light Source, a national research facility of University of Saskatchewan, which is supported by CFI, NSERC, the National Research Council (NRC), the Canadian Institutes of Health Research (CIHR), the Government of Saskatchewan, and the University of Saskatchewan. The STEM work at the Brookhaven National Laboratory was supported by the U.S. DOE Basic Energy Sciences, Materials Science and Engineering Division under Contract No. DESC0012704. This research used the focused ion beam instrument at the Center for Functional Nanomaterials, a U.S. Department of Energy Office of Science User Facility, at the Brookhaven National Laboratory. C.H. and O.D. were financially supported by Intellectual Ventures - Deep Science Fund. C.H. also acknowledges the Austrian Science Fund (FWF) Project No. P 32144-N36.

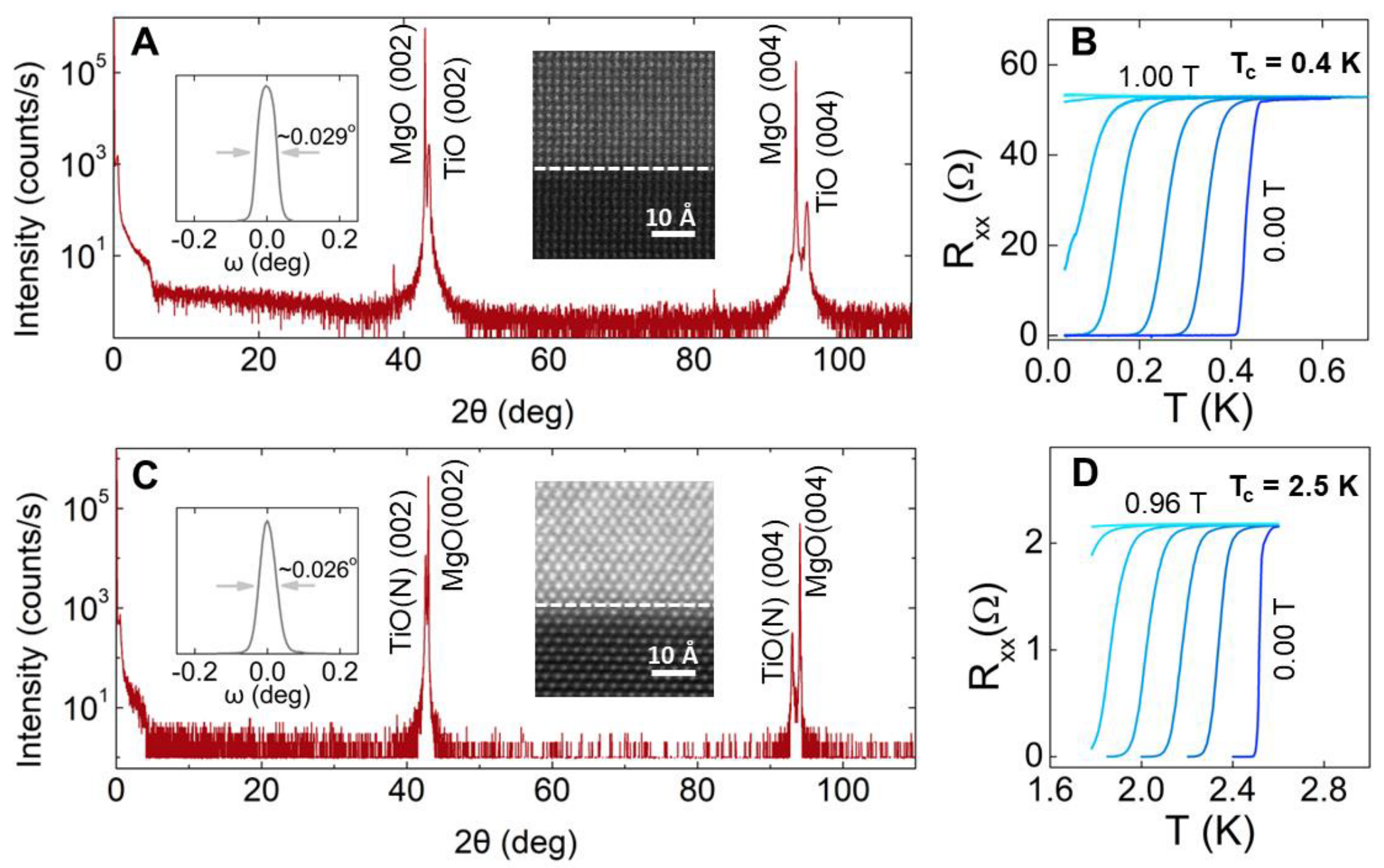

**Fig. 1. Coherent growth of crystalline TiO and TiO(N) epitaxial films on MgO(001) substrates.** (**A**) and (**C**) the XRD 2θ-ω scans of TiO and TiO(N) crystalline structure, respectively. Insets (left): rocking curves at the (002) diffraction of films; Inset (right): the HADDF images measured in the STEM experiments. The probe electron beams are parallelled to the [100] and [111] directions for TiO (**A**) and TiO(N) (**B**), respectively. The green dashed lines indicate the film and substrate interfaces. (**B**) and (**D**) the superconductivity transition and magnetoresistance of TiO and TiO(N) films, respectively. The superconductivity $T_c$ is suppressed by applying out-of-the-plane magnetic fields.

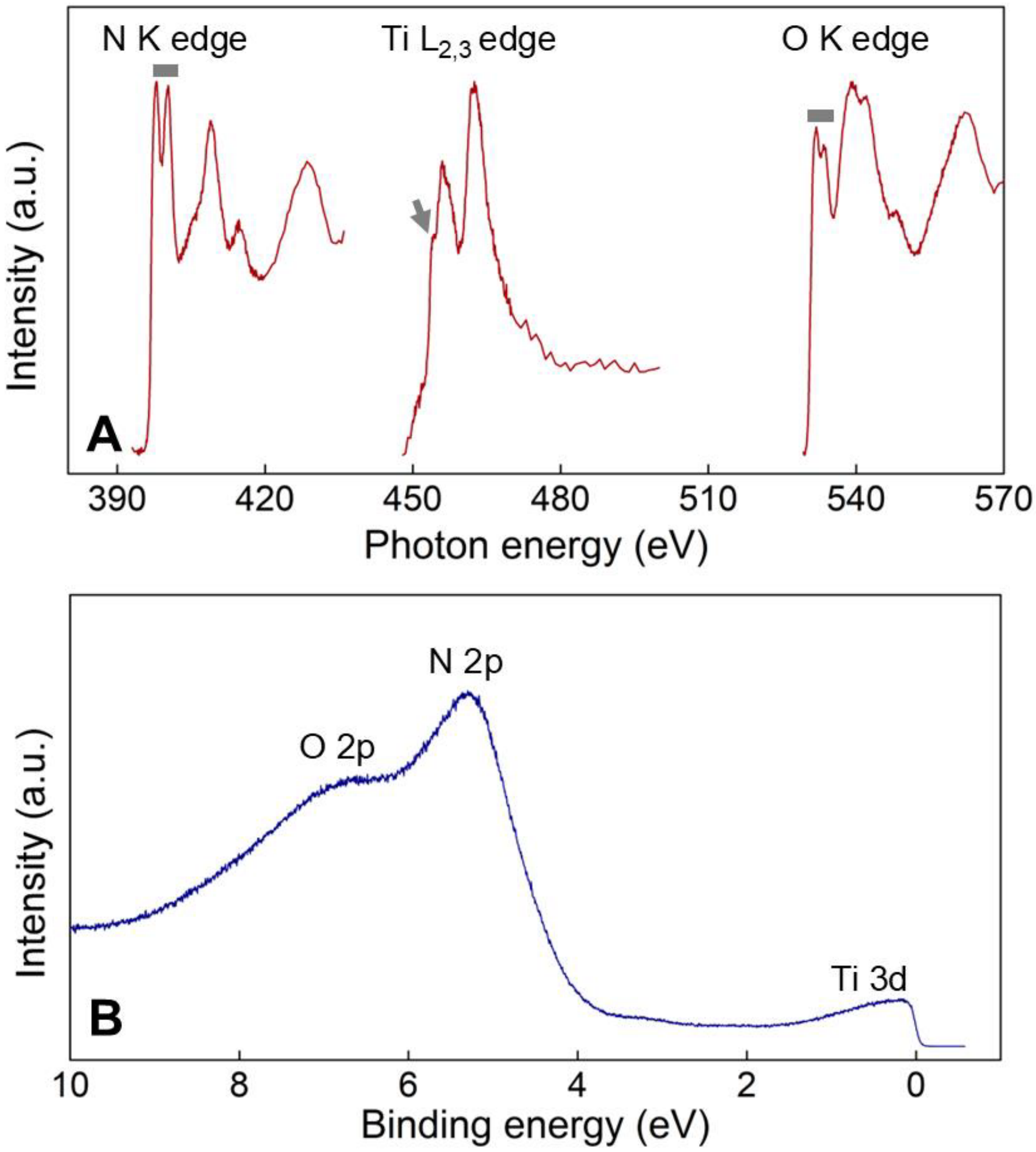

**Fig. 2. Spectroscopy characterization of TiO(N) films.** (**A**) X-ray absorption spectroscopy (XAS) at the N K, Ti $L_{2,3}$ and O K edges (From left to right). The horizontal solid lines just above the O and N spectra underscore double peaks due to the excitation of 1s electrons in O and N to their unoccupied p orbitals that hybridize strongly with Ti electronic states. The arrow indicating the shoulder at Ti $L_{2,3}$ edge results from N substitution induced $Ti^{3+}$. (**B**) The occupied N and O 2p states, and Ti 3d $t_{2g}$ states measured using i*n-situ* ultraviolet photoemission spectroscopy with the photon energy at 21.2 eV. The Fermi level is set at E=0.

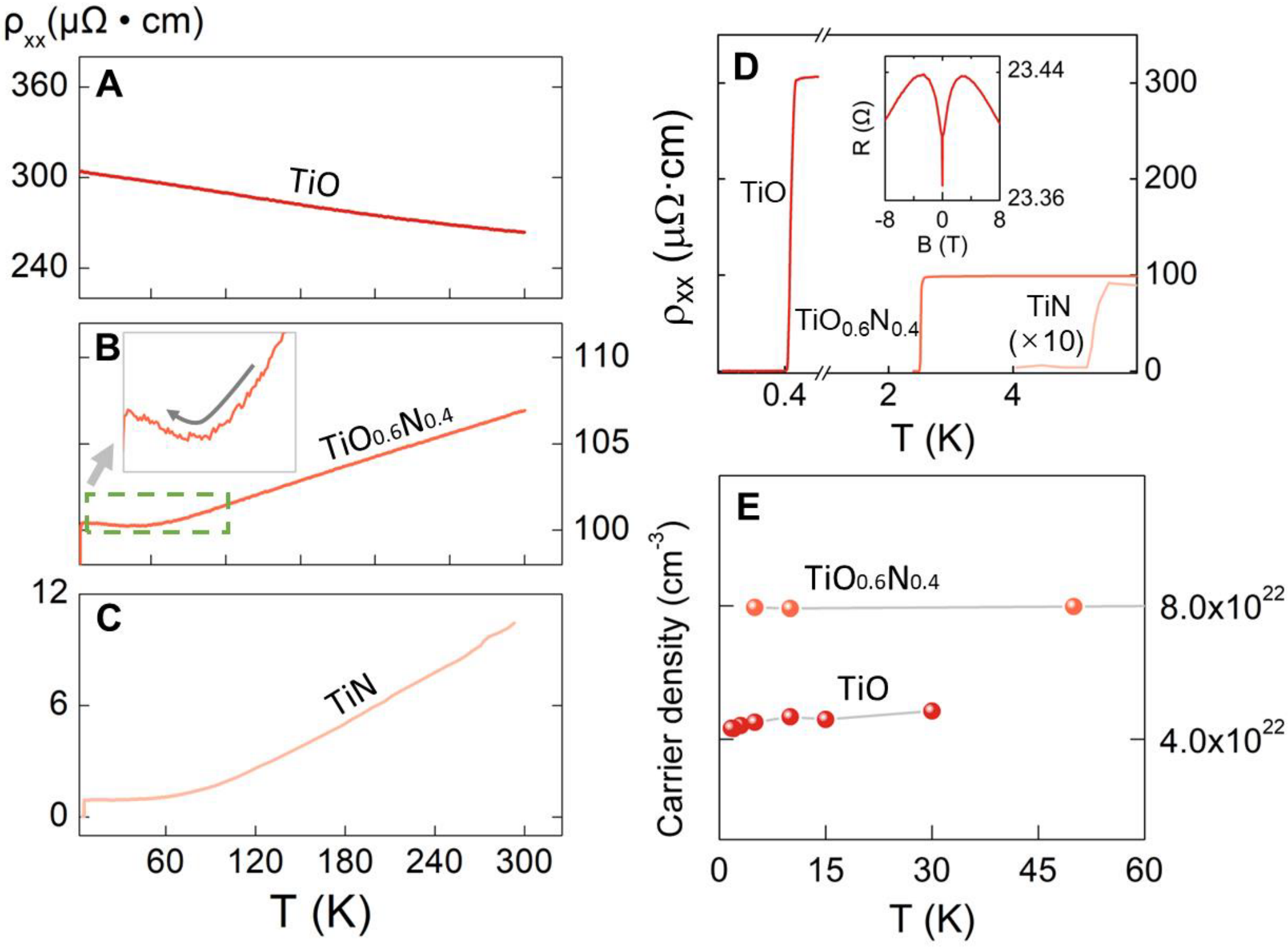

**Fig. 3.** Electrical transport results of epitaxial films in the Ti(O,N) series. (**A-C**) Resistivity versus temperature of TiO (top), $TiO_{0.6}N_{0.4}$ (middle) and TiN (bottom, the results adapted from Ref. (*14*). Inset in (B) highlights the small upturn of resistivity near 60 K as the sample is cooled down. (**D**) Superconducting transition temperatures ($T_c$) are lowered as the normal-state resistivity increases. Inset: The TiO magnetoresistance at 2 K. (**E**) Carrier density in TiO and $TiO_{0.6}N_{0.4}$ at different temperatures from the Hall measurements. The type of carrier in $TiO_{0.6}N_{0.4}$ is electron, while TiO has hole-type itinerant carriers.

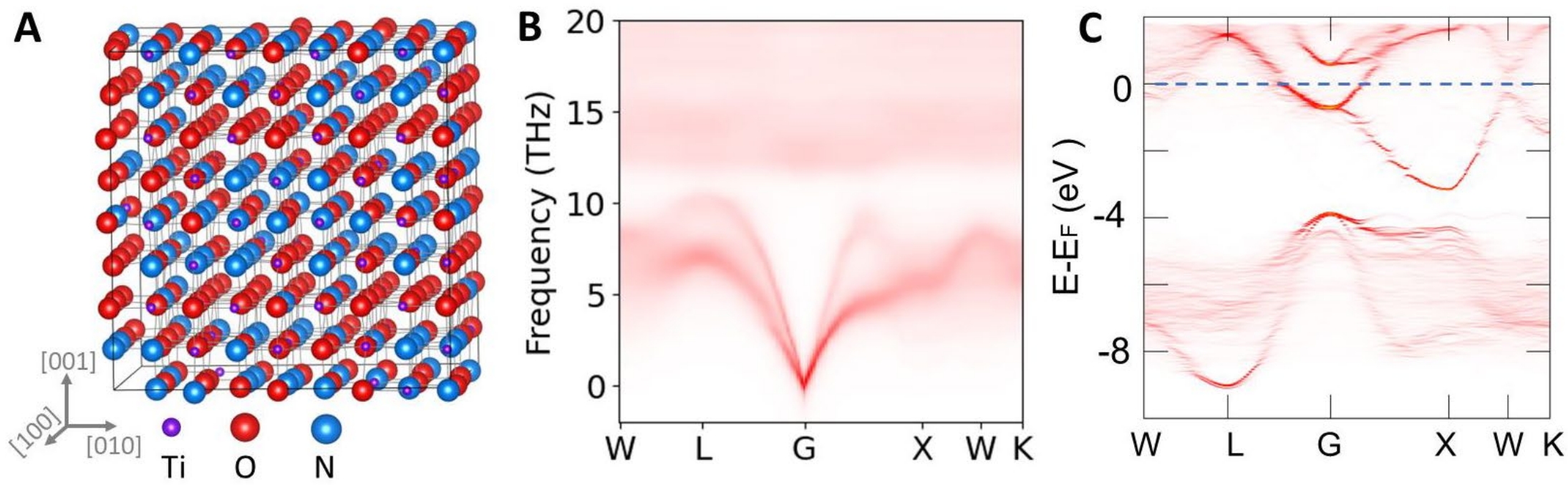

**Fig. 4.** Disorder effects on the band and phonon structures. (**A**) The structurally relaxed supercell with quasi-randomly distributed N and O at anion sites, *i.e.*, a nominal stoichiometry $TiO_{0.61}N0_{.39}$, is used for calculating the unfolded phonon (**B**) and band dispersions (**C**). The color weight indicates the coherence of the Bloch wavefunctions (*35*).

# Supplementary Materials for

## Critical Role of Disorder for Superconductivity in the Series of Epitaxial Ti(O,N) Films

**Authors:** Fengmiao Li[1,2], Oliver Dicks[2], Myung-Geun Han[3], Solveig Aamlid[2], Giorgio Levy[1,2], Ronny Sutarto[4], Chong Liu[1,2], Hsiang-Hsi Kung[1,2], Oleksandr Foyevstov[1,2], Simon Godin[1,2], Bruce A. Davidson[1,2], Andrea Damascelli[1,2], Yimei Zhu[3], Christoph Heil[5], Ilya Elfimov[1,2], George A. Sawatzky[1,2], Ke Zou[1,2]

**Affiliations:**

[1]Department of Physics & Astronomy, University of British Columbia, Vancouver, British Columbia, V6T 1Z1 Canada.

[2]Quantum Matter Institute, University of British Columbia, Vancouver, British Columbia, V6T 1Z4 Canada.

[3]Condensed Matter Physics & Materials Science Department, Brookhaven National Laboratory, Upton, New York, 11973 USA.

[4]Canadian Light Source, Saskatoon, Saskatchewan, S7N 2V3 Canada.

[5]Institute of Theoretical and Computational Physics, Graz University of Technology, NAWI Graz, 8010 Graz, Austria

## This PDF file includes:

## Figs. S1-S8

S1 Lattice structural characterization using electron and x-ray diffractions.
S2 Resistance vs magnetic field at temperatures near the Tc.
S3 X-ray photoemission spectroscopy (XPS) of uncapped TiO(N) films.
S4 Electron energy loss spectroscopy (EELS) of TiO(N) films in STEM.
S5 Ultraviolet photoemission spectroscopy (UPS) and x-ray absorption spectroscopy (XAS) measured on TiO(N) films.
S6 Total density of states of TiO, $TiO_{0.61}N_{0.39}$ and TiN.
S7 The calculated phonon dispersion of the bulk TiO and TiN.
S8 Band structures of TiO, $TiO_{0.61}N_{0.39}$ and TiN.

**Figure S1**

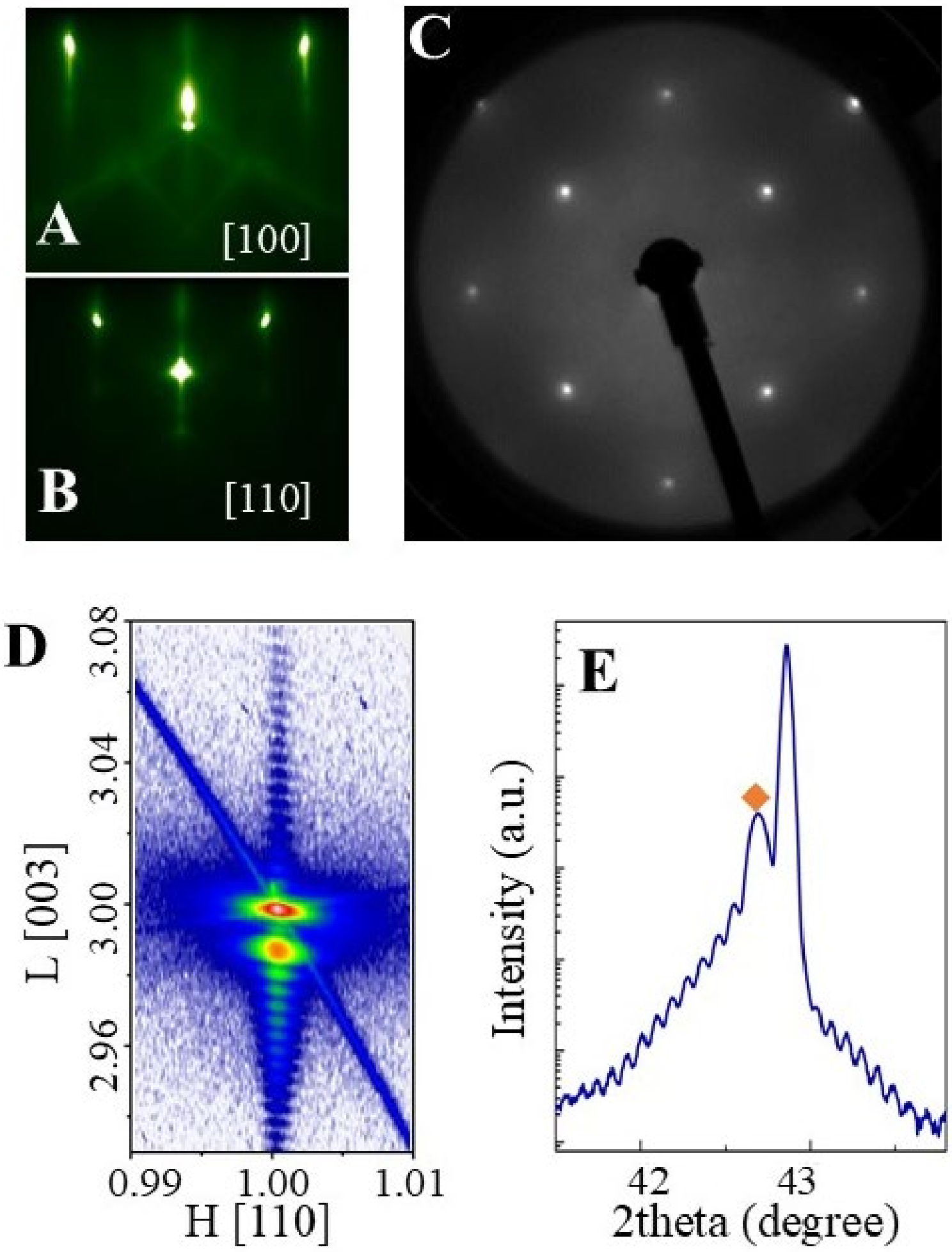

**Figure. S1. Lattice structural characterization using electron and x-ray diffractions.** (**A**) and (**B**) RHEED patterns of a 90 nm film surface with the incident electron beam along [100] and [110] direction, respectively. (**C**) the LEED pattern of the film surface collected using a 180eV electron beam. (**D**) the reciprocal space mapping near the MgO and film (113) diffraction region. (**E**) XRD 2θ-ω scan around the (002) diffraction and. The diamond symbol indicates the 002-diffraction peak of the film with the thickness fringes showing the high crystalline quality.

## Figure S2

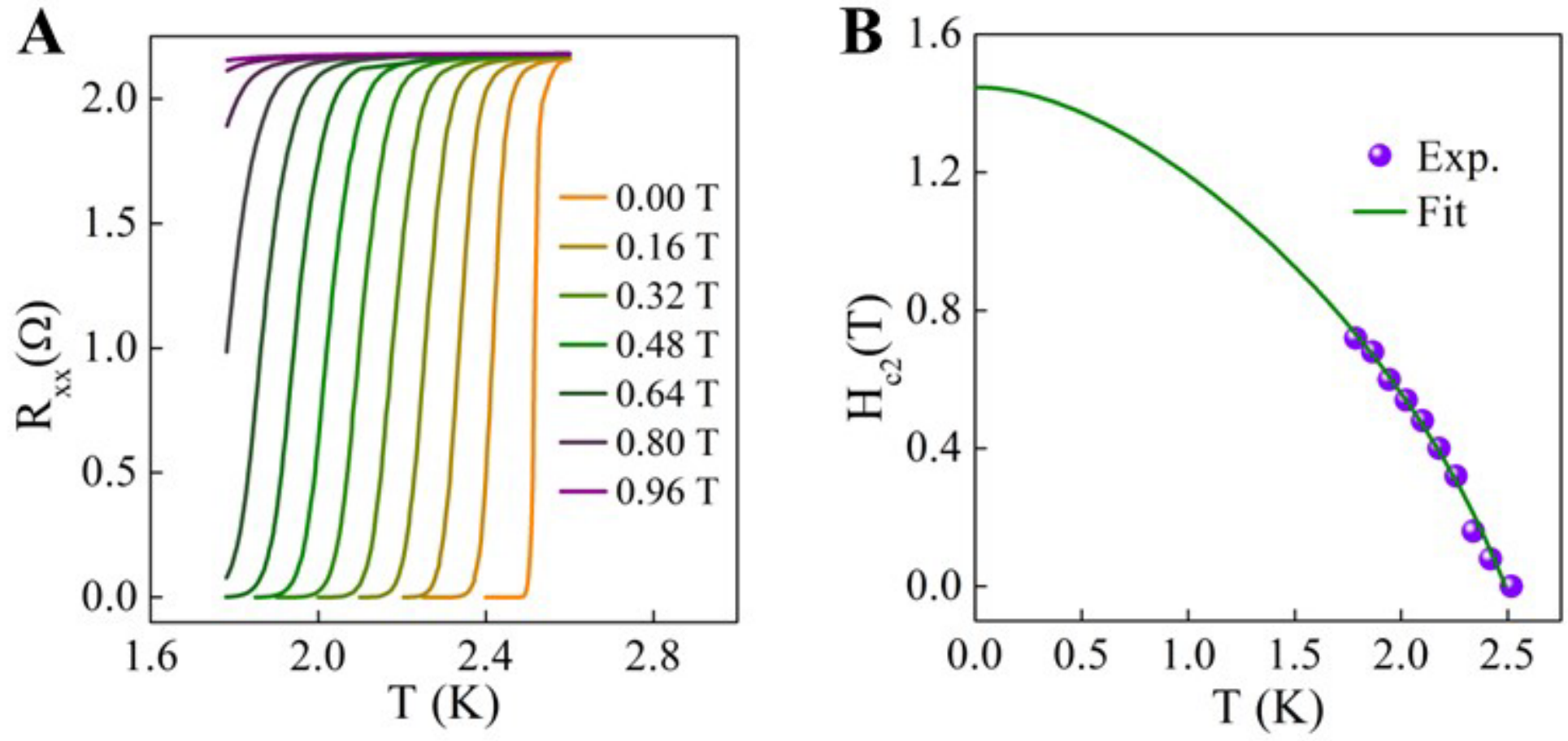

**Figure. S2. Resistance vs magnetic field at temperatures near the $T_c$.** (**A**) Longitudinal resistance $R_{xx}(T)$ as a function of magnetic field perpendicular to film surface with T < 2.5 K. The upper critical field $H_{c2}$ (in Tesla) is obtained using 50% criterion of the normal-state resistance value in (**A**). $H_{c2}(0)$ is extracted from the fitting (**B**) of $H_{c2}$ (in Tesla) with the Werthamer-Helfand-Hohenberg (WHH) mode (N. R. Werthamer et al., Phys. Rev. 147, 295–302 (1966)).

## Figure S3

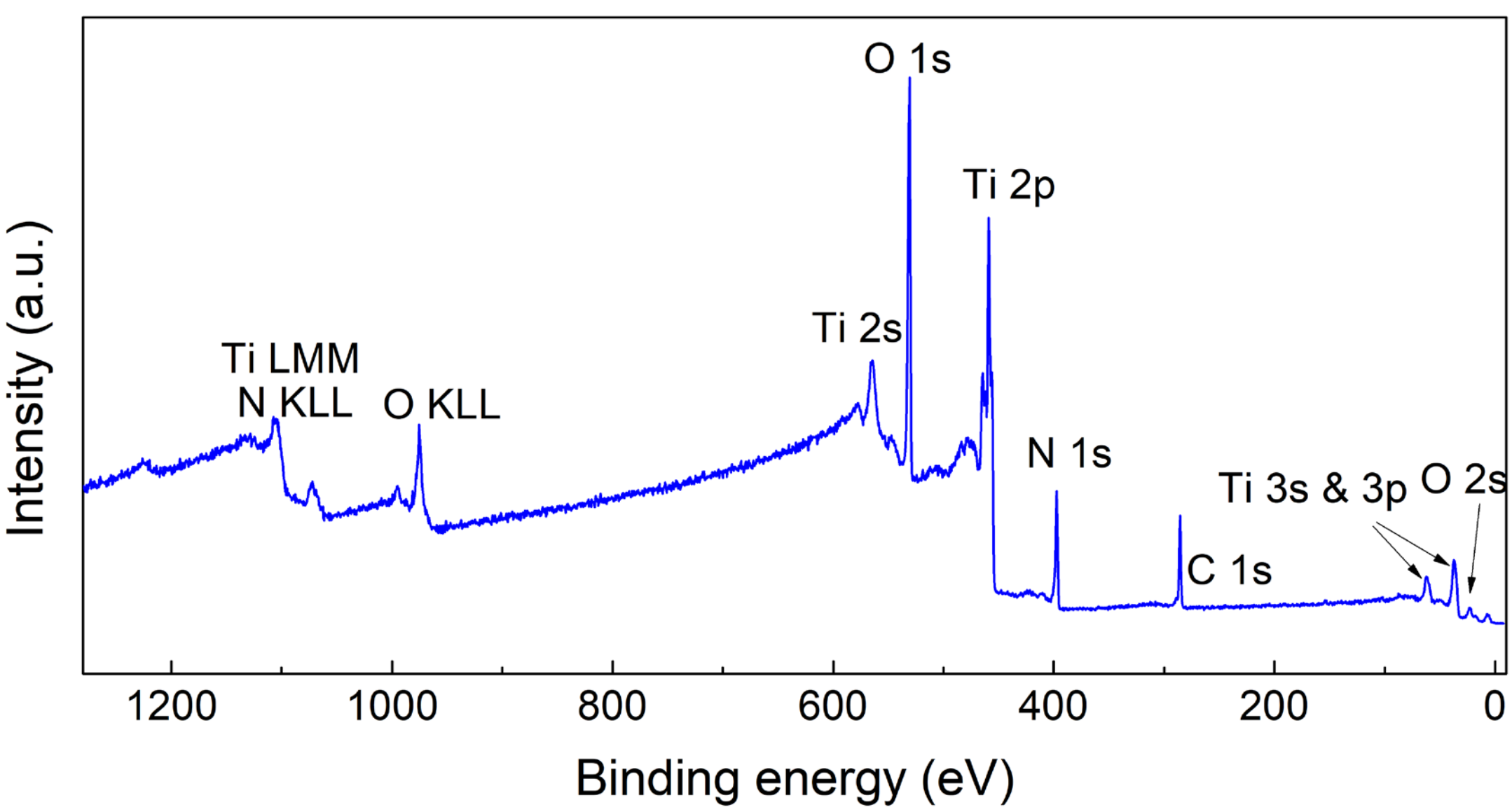

**Figure. S3. X-ray photoemission spectroscopy (XPS) of uncapped TiO(N) films.** Carbon and stronger oxygen signal are seen in the spectroscopy, caused by surface contamination and oxidation after air exposure.

## Figure S4

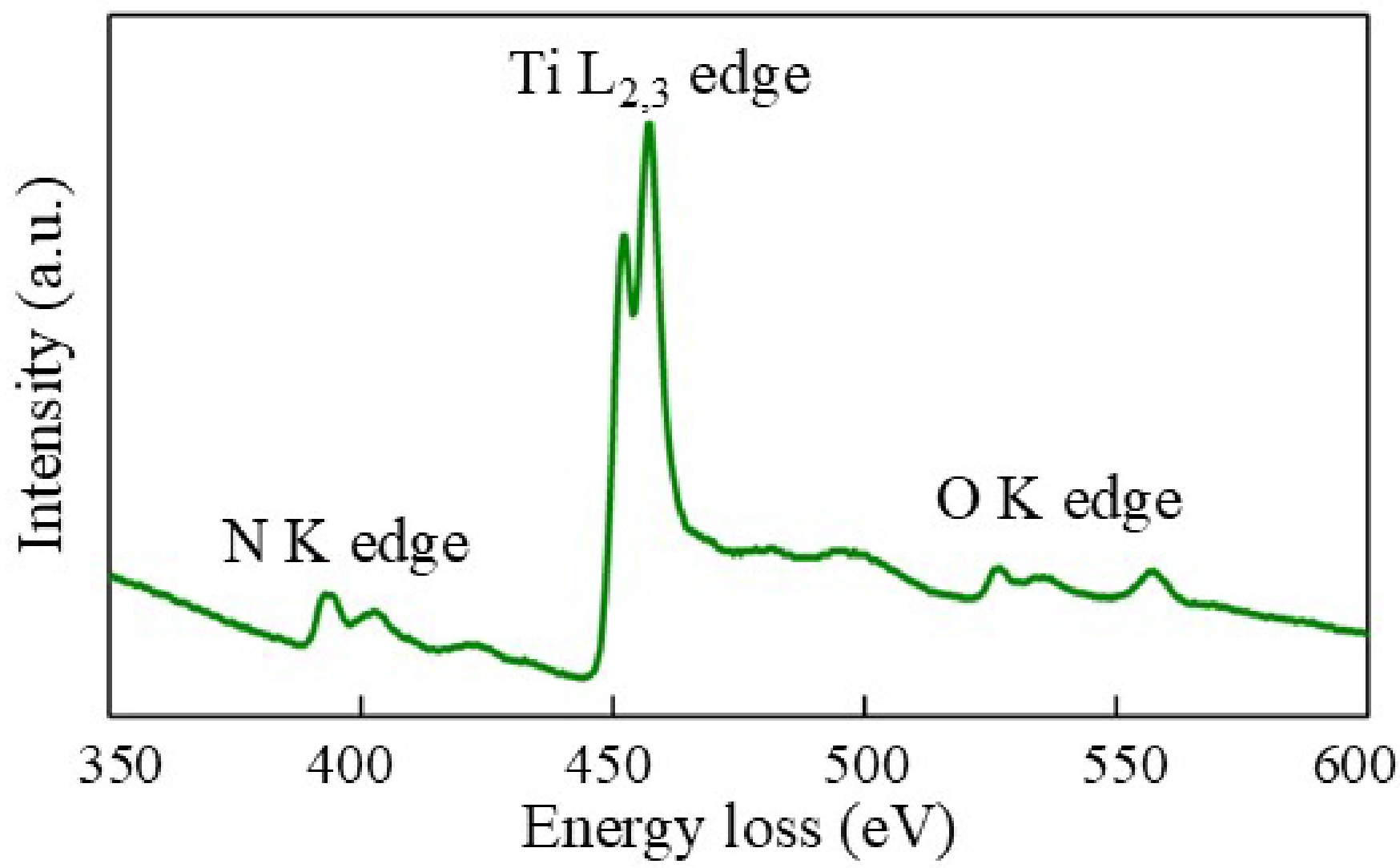

**Figure. S4. Electron energy loss spectroscopy (EELS) of TiO(N) films in STEM.** The O K edge, Ti $L_{2,3}$ edge and N K edge are visible in the energy range between 350 eV and 600 eV.

**Figure S5**

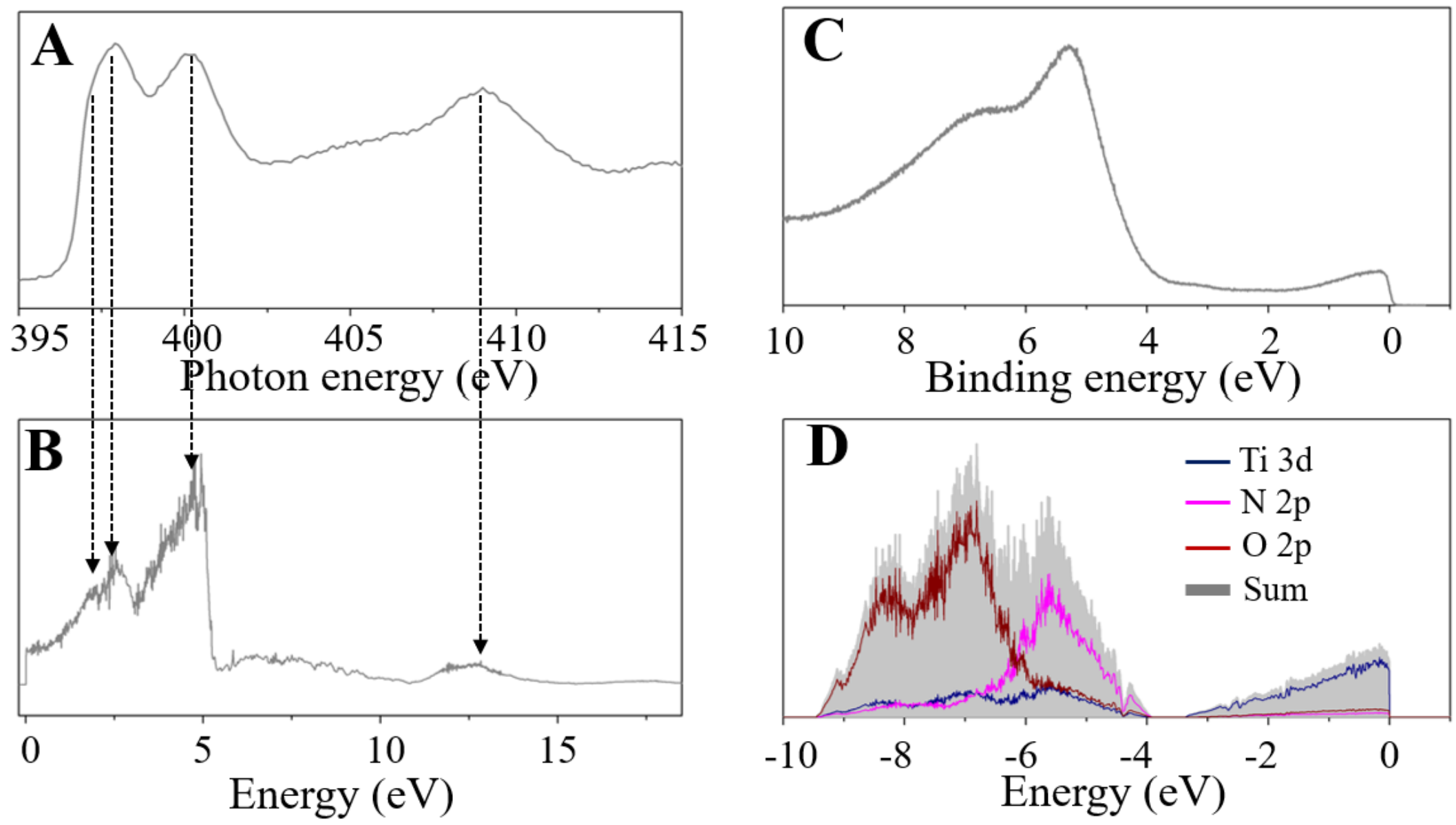

**Figure. S5. Ultraviolet photoemission spectroscopy (UPS) and x-ray absorption spectroscopy (XAS) measured on TiO(N) films.** The XAS at the N K edge (**A**) and the projected p state of the N (**B**) calculated from the DFT calculation using a supercell containing 64 Ti, 25 N and 39 O. The arrows are used to guide the eye for the comparison. The UPS (**C**) compares with the calculated spectrum from the DFT DOS weighted by the cross section of each atomic orbital (**D**).

## Figure S6

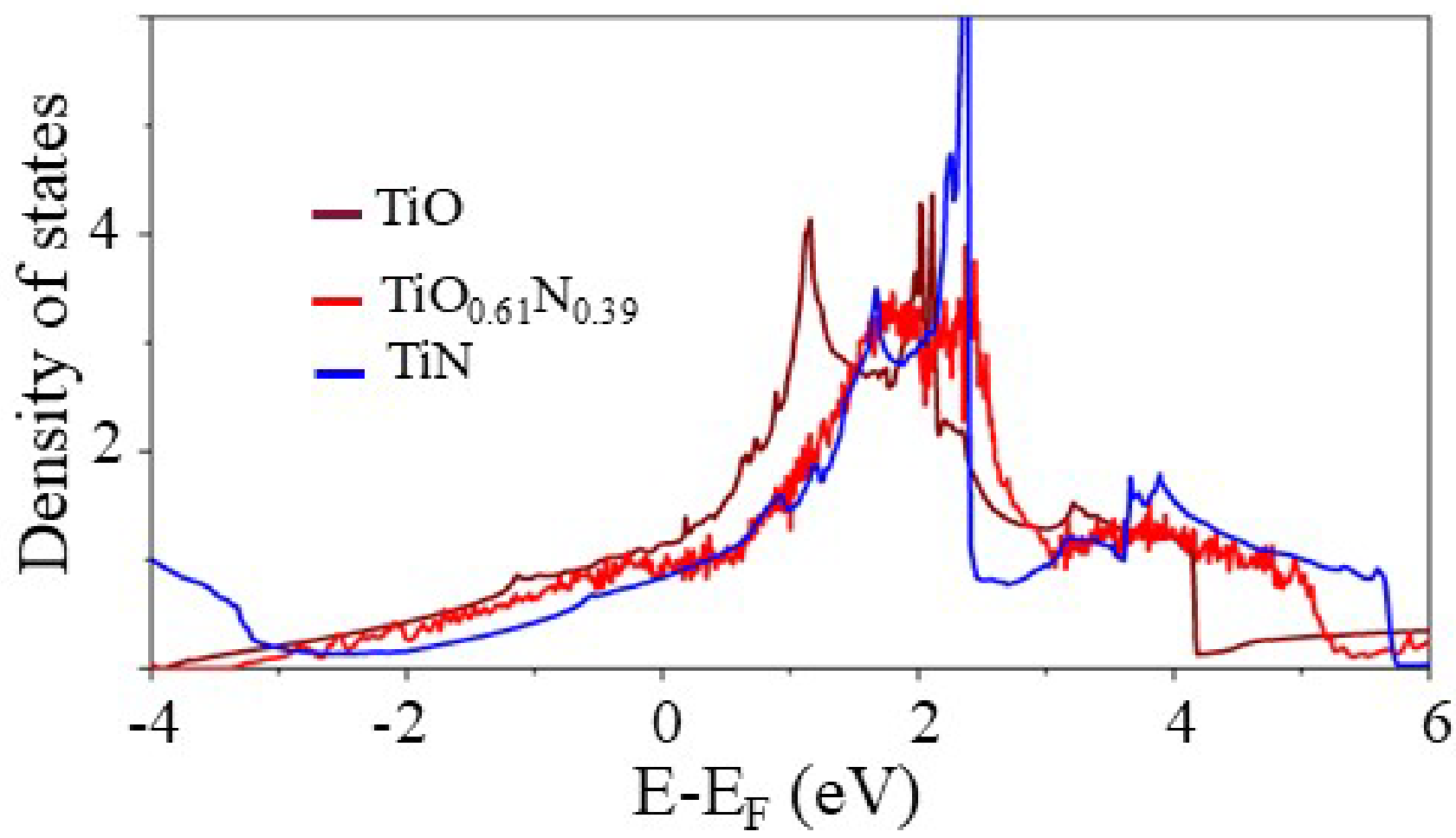

**Figure. S6. Total density of states of TiO, $TiO_{0.61}N_{0.39}$ and TiN.** The density of states (DOS) unit is states/eV/TiO (TiN or $TiO_{0.61}N_{0.39}$).

## Figure S7

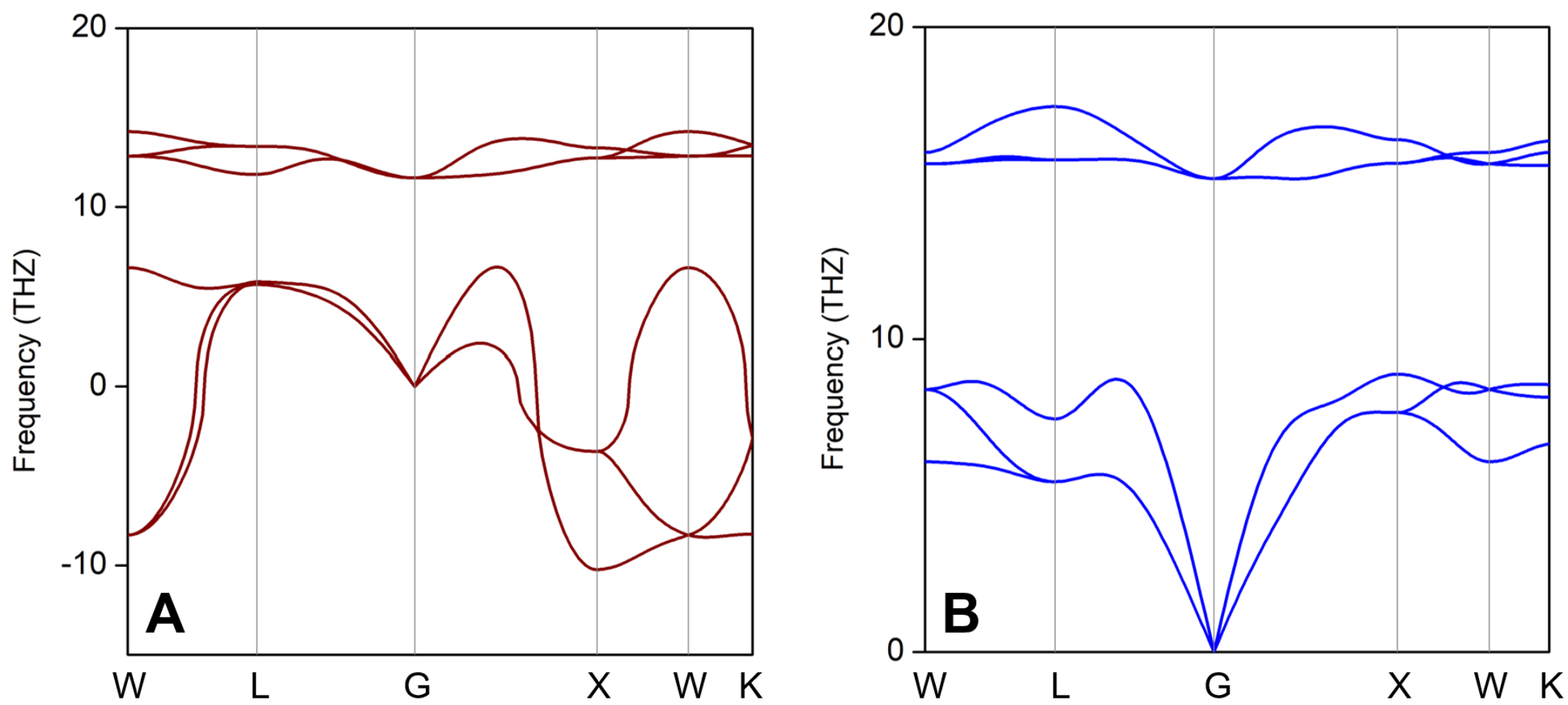

**Figure. S7. The calculated phonon dispersion of the bulk TiO (A) and TiN (B)**. The substantial phonons with negative frequencies suggest the intrinsic instability of TiO with the high-symmetry Rocksalt atomic structure.

## Figure S8

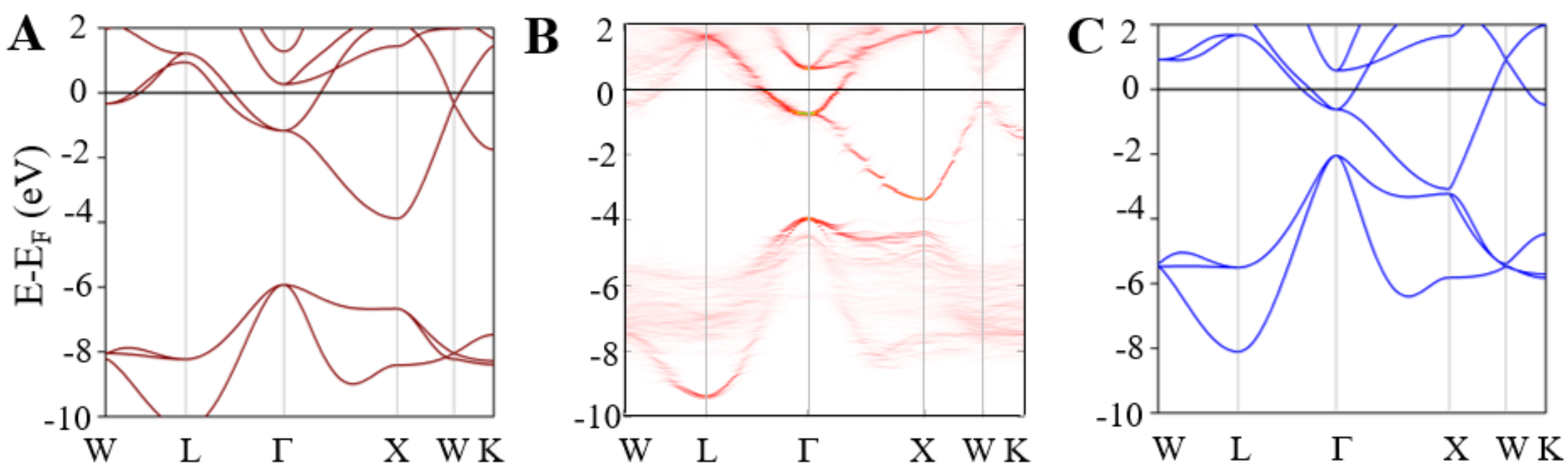

**Figure. S8. Band structures of TiO, $TiO_{0.61}N_{0.39}$ and TiN.** The low-energy band dispersion of O 2p and Ti 3d $t_{2g}$ in the bulk and Rocksalt TiO (**A**) and TiN (**C**), and structurally relaxed $TiO_{0.61}N_{0.39}$ (**B**). The unfolded band dispersion in (**B**) is calculated from a Ti oxynitride supercell. The E = 0 eV is set as the Fermi energy.